\documentclass[conference]{IEEEtran}
\IEEEoverridecommandlockouts
\usepackage{cite}
\usepackage{amsmath,amssymb,amsfonts}
\usepackage{algorithmic}
\usepackage{graphicx}
\usepackage{textcomp}
\usepackage{xcolor}
\usepackage{times}
\usepackage{helvet}
\usepackage{courier}
\usepackage{bm}
\usepackage{times}
\usepackage{soul}
\usepackage{url}
\usepackage[utf8]{inputenc}
\usepackage[small]{caption}
\usepackage{graphicx}
\usepackage{amsmath}
\usepackage{amsthm}
\usepackage{booktabs}
\usepackage{algorithm}
\usepackage{algorithmic}
\usepackage[switch]{lineno}
\def\BibTeX{{\rm B\kern-.05em{\sc i\kern-.025em b}\kern-.08em
    T\kern-.1667em\lower.7ex\hbox{E}\kern-.125emX}}
\begin{document}

\title{Enhanced Creativity and Ideation through Stable Video Synthesis\\
}

\author{\IEEEauthorblockN{Elijah Miller, Thomas Dupont, Mingming Wang}
USC, wangmm913@gmail.com \\

}

\twocolumn[{
\renewcommand\twocolumn[1][]{#1}
\maketitle
\begin{center}
    \captionsetup{type=figure}
    	\includegraphics[width=0.95\linewidth]{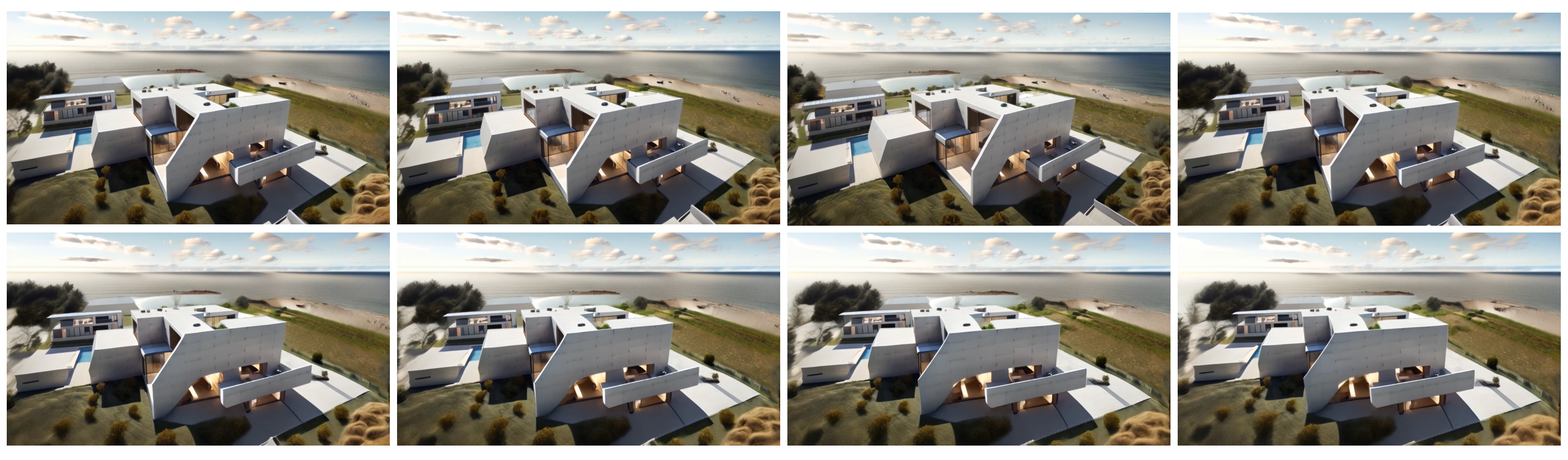}
	\caption{These sequences showcase the innovative application of Stable Video Diffusion (SVD), where a static landscape image is transformed into a dynamic, evolving scenery, resulting in visually consistent and natural movements that bring the landscape to life.}
	\label{fig:teaser}
\end{center}
}]

\begin{abstract}
This paper explores the innovative application of Stable Video Diffusion (SVD), a diffusion model that revolutionizes the creation of dynamic video content from static images. As digital media and design industries accelerate, SVD emerges as a powerful generative tool that enhances productivity and introduces novel creative possibilities. The paper examines the technical underpinnings of diffusion models, their practical effectiveness, and potential future developments, particularly in the context of video generation.
SVD operates on a probabilistic framework, employing a gradual denoising process to transform random noise into coherent video frames. It addresses the challenges of visual consistency, natural movement, and stylistic reflection in generated videos, showcasing high generalization capabilities. The integration of SVD in design tasks promises enhanced creativity, rapid prototyping, and significant time and cost efficiencies. It is particularly impactful in areas requiring frame-to-frame consistency, natural motion capture, and creative diversity, such as animation, visual effects, advertising, and educational content creation. The paper concludes that SVD is a catalyst for design innovation, offering a wide array of applications and a promising avenue for future research and development in the field of digital media and design.
\end{abstract}


\section{Introduction}

In the field of digital media and design, the creation and processing of visual content are advancing at an unprecedented pace. Particularly in the realm of video generation~\cite{videocontrolnet,videoLDM,blattmann2023stable}, technological advancements have not only enhanced creators' productivity but also introduced new creative possibilities. Generating videos from static images is a significant technology that transforms static artworks, photos, or other image materials into dynamic video content, enriching visual expression. Recently, diffusion models~\cite{rombach2022high,imagen} have emerged as powerful generative models, demonstrating substantial potential in both image and video generation~\cite{magicvideo,wang2018video2video,ceylan2023pix2video}. This paper aims to explore the application of diffusion models in the task of generating videos from static images, analyzing their technical principles, practical effectiveness, and future development directions.

Diffusion models are a class of probabilistic models that generate data through a gradual denoising process, inspired by diffusion processes in physics and widely applied in recent deep learning research. These models define a reverse process from a simple distribution (like Gaussian noise) to the target data distribution. The key to this generation method is that the model learns a series of denoising steps, transforming random noise into clear images or video frames step by step. Generating videos from static images involves expanding a single static image into a coherent and natural sequence of frames, a task fraught with challenges. Firstly, the generated video must maintain visual consistency, avoiding noticeable distortions or artifacts. Secondly, the video generation needs to capture natural movement and change, which is crucial for maintaining dynamic effects. Furthermore, the generated videos must reflect the unique style and content characteristics of different static images, demanding high generalization capabilities from the model.

Diffusion models have emerged as particularly successful, swiftly eclipsing methods based on generative adversarial networks (GANs) and auto-regressive Transformers in becoming the predominant approach for image generation. Notable works in this domain include Stable Diffusion \cite{stablediffusion}, DALL-E \cite{dalle2}, Midjourney \cite{Midjourney}, ControlNet \cite{controlnet}, DreamBooth \cite{ruiz2023dreambooth}, Cascaded Diffusion \cite{cascaded}, and Imagen \cite{imagen}.

Diffusion-based methods are favored for their strong controllability, ability to produce photorealistic images, and remarkable diversity. Moreover, these techniques find applications across a wide spectrum of computer vision tasks, including image editing \cite{brooks2023instructpix2pix,hertz2022prompt2prompt,li2023layerdiffusion,li2024tuning,avrahami2022blended,tumanyan2023plug}, dense prediction \cite{diffusiondet,li2023efficient,diffmatch}, as well as in domains such as video synthesis \cite{singer2022make,vdm,ho2022imagenvideo,videofusion,videofactory,SimDA} and 3D generation \cite{magic3d,dreamfusion,diffusion3d,li2023archi,li2024generating,text-to-3d,lin2023magic3d,raj2023dreambooth3d,li2024art3d}.

Recent research in the field of video generation has made significant strides. Notable works include Denoising Diffusion Probabilistic Models (DDPM)~\cite{ho2020denoising,improvedddpm,sohl2015deep}, which achieved high-quality image generation through a gradual denoising process and demonstrated potential in video generation. Score-based Generative Models by Song et al. leveraged score matching to learn the gradient information of data distribution, enabling efficient sampling in high-dimensional data spaces. These studies not only validate the effectiveness of diffusion models in generative tasks but also provide theoretical foundations and technical support for further exploration of their applications in practical design tasks.

The integration of Stable Video Diffusion (SVD)~\cite{blattmann2023stable} in the design process has far-reaching implications. It enhances creativity by providing designers with a plethora of video concepts that they might not have initially considered, leading to innovative designs that push the boundaries of traditional design thinking. The rapid prototyping capabilities of SVD accelerate the design process, allowing for quick testing of different design ideas without the need for extensive post-production work. This results in significant time and cost savings, making high-quality video production accessible to a broader range of designers.

In video generation, diffusion models are primarily applied in several key areas: frame-to-frame consistency, capturing natural motion, and diversity and creativity. By using a gradual denoising process, diffusion models can maintain high consistency between generated frames, helping to avoid common issues like frame flickering and discontinuity seen in traditional generation methods. Through learning a series of complex denoising steps, diffusion models can capture the natural movement patterns, generating smooth and realistic video sequences, which makes them particularly advantageous in animation and visual effects production. Additionally, diffusion models exhibit strong generative diversity, capable of producing various dynamic videos from the same static image, offering creators a rich creative space and more design options.

In practical design tasks, the application scenarios of generating videos from static images using SVD are vast and diverse. As shown in Fig. 1, artists can transform static artworks into dynamic videos, enhancing their expressiveness and viewer appeal. Businesses can create more engaging advertising content from product images, improving brand promotion effects. In game and film production, SVD can be used to generate cutscenes and special effects, saving on production time and costs. Educational institutions can leverage SVD to create vivid teaching materials, enhancing the learning experience.

\section{Formulas and Mathematical Framework}

In recent years, diffusion models~\cite{ho2020denoising,song2020score} have gained significant attention in the field of generative modeling. Current research on diffusion models is predominantly centered on three formulations: denoising diffusion probabilistic models (DDPMs)~\cite{ho2020denoising,improvedddpm}, score-based generative models (SGMs)~\cite{song2019generative,song2020improved}, and stochastic differential equations (Score SDEs)~\cite{song2021maximum}.
This section delves into the mathematical principles and formulas underlying diffusion models.

\paragraph{Forward Diffusion Process}The forward diffusion process involves gradually adding noise to the data, transforming it step by step into a noisy version that approximates a simple distribution, typically Gaussian noise. Let \( \mathbf{x}_0 \) be an initial data point (e.g., an image). The forward process defines a sequence of latent variables \( \mathbf{x}_1, \mathbf{x}_2, \ldots, \mathbf{x}_T \), where \( \mathbf{x}_T \) is nearly pure noise. This process can be described as:
\begin{equation}
q(\mathbf{x}_t | \mathbf{x}_{t-1}) = \mathcal{N}(\mathbf{x}_t; \sqrt{\alpha_t} \mathbf{x}_{t-1}, (1 - \alpha_t) \mathbf{I})     
\end{equation}

Here, \( \alpha_t \) is a parameter associated with the time step \( t \), typically chosen such that \( 0 < \alpha_t < 1 \).

The overall forward diffusion process can be represented as a joint distribution:
\begin{equation}
q(\mathbf{x}_{1:T} | \mathbf{x}_0) = \prod_{t=1}^{T} q(\mathbf{x}_t | \mathbf{x}_{t-1})
\end{equation}
At the final time step \( T \), the distribution of \( \mathbf{x}_T \) should ideally be close to a standard normal distribution:
\begin{equation}
q(\mathbf{x}_T | \mathbf{x}_0) = \mathcal{N}(\mathbf{x}_T; \sqrt{\bar{\alpha}_T} \mathbf{x}_0, (1 - \bar{\alpha}_T) \mathbf{I})
\end{equation}
where \( \bar{\alpha}_T = \prod_{t=1}^{T} \alpha_t \).

\paragraph{Reverse Diffusion Process}

The reverse diffusion process aims to recover the original data \( \mathbf{x}_0 \) from the noisy variable \( \mathbf{x}_T \) by gradually removing the noise. The reverse process is modeled by learning a parameterized distribution \( p_\theta \) that approximates the true reverse of the forward process. Each step of the reverse process can be written as:
\begin{equation}
p_\theta(\mathbf{x}_{t-1} | \mathbf{x}_t) = \mathcal{N}(\mathbf{x}_{t-1}; \mu_\theta(\mathbf{x}_t, t), \Sigma_\theta(\mathbf{x}_t, t))
\end{equation}

where \( \mu_\theta \) and \( \Sigma_\theta \) are functions parameterized by a neural network, which need to be learned from data.

The joint distribution of the reverse process is given by:
\begin{equation}
p_\theta({x}_{0:T}) = p({x}_T) \prod_{t=1}^{T} p_\theta({x}_{t-1} | {x}_t)
\end{equation}

Here, \( p({x}_T) \) is typically assumed to be a standard normal distribution:
\begin{equation}
p({x}_T) = \mathcal{N}({x}_T; {0}, {I})
\end{equation}

\paragraph{Training Objective} To train the reverse diffusion model, we maximize the variational lower bound (VLB) on the data likelihood. This involves minimizing the Kullback-Leibler (KL) divergence between the true reverse process and the learned reverse process:
\begin{equation}
{E}_{q} \left[ \log p_\theta({x}_0) \right] \geq {E}_{q} \left[ \log \frac{p_\theta(f{x}_{0:T})}{q({x}_{1:T} | \mathbf{x}_0)} \right]
\end{equation}

This objective can be decomposed into a sum of per-step reconstruction errors:
\begin{equation}
L_{\text{vlb}} = {E}_{q} \left[ \sum_{t=1}^{T} \text{KL} \left[A\right]- \log p_\theta({x}_0 | {x}_1) \right]
\end{equation}
\begin{equation}
A =   q({x}_{t-1} | {x}_t, {x}_0) \| p_\theta({x}_{t-1} | {x}_t) 
\end{equation}

In practical implementations, such as in Denoising Diffusion Probabilistic Models (DDPM), a simplified objective is often optimized. This simplified objective focuses on denoising score matching, where the model learns to predict the added noise at each step:
\begin{equation*}
L_{\text{simple}} = {E}_{{x}_0, \boldsymbol{\epsilon}, t} \left[ \| \boldsymbol{\epsilon} - \boldsymbol{\epsilon}_\theta({x}_t, t) \|^2 \right]
\end{equation*}
Here, \( \boldsymbol{\epsilon} \) is the noise sampled from a standard normal distribution, and \( \boldsymbol{\epsilon}_\theta \) is the neural network's prediction of the noise.

\section{Stable Video Diffusion}

Stable Video Diffusion (SVD)~\cite{blattmann2023stable} is a model for generating high-resolution videos based on text or image prompts. It leverages the concept of latent diffusion models, which have shown success in image synthesis and are now adapted for video generation. It highlight the importance of training strategies and data selection in the performance of generative video models.

\subsection{Model Architecture} The SVD model is built upon a latent space approach, reducing computational complexity. It incorporates temporal layers into a pretrained text-to-image model. As shown in Fig. 2, the architecture follows the design from~\cite{blattmann2023stable}, with temporal convolution and attention layers inserted after every spatial layer. The model is trained end-to-end, unlike some approaches that only train temporal layers.

\subsection{Training Strategies}

SVD identify three critical stages for effective training of video latent diffusion models (LDMs). Text-to-Image Pretraining: Starting with a strong foundation in image generation. Video Pretraining: Training on a large dataset at a lower resolution to learn general motion representations. High-Quality Video Finetuning: Refining the model on a smaller, high-quality dataset for improved detail and resolution.

\subsection{Data Curation} A significant portion of the paper is dedicated to the systematic curation of video data. SVD propose a method to transform a vast, unfiltered collection into a curated dataset suitable for training generative video models. The curation process includes: Cut Detection: Removing clips with cuts or fades to improve the training quality. Annotation: Using synthetic captioning methods and optical flow to assess motion and aesthetics. Filtering: Applying thresholds to remove static scenes, excessive text, and low-quality clips.

\subsection{Training Procedure}

The training process involves several steps: Image Pretraining: Utilizing a pretrained image diffusion model to provide a strong visual representation. Video Pretraining: Training on a large, curated dataset to learn motion and video-specific features. Finetuning: Refining the model on high-quality videos to achieve state-of-the-art results. The paper detail the use of a continuous noise schedule and network preconditioning from Karras et al. to adapt the model for higher resolutions.

\subsection{Implementation Details}
The paper provides a detailed account of the model's implementation: UNet Architecture: The use of a UNet structure for spatial feature extraction, extended with temporal layers. Temporal Layers: Implementation of temporal convolution and attention mechanisms. Noise Schedule: Shifting the noise schedule towards higher values for better high-resolution results. Classifier-Free Guidance: Employing a method to guide the diffusion process towards a conditioning signal.

\begin{figure}[t]
	\centering

	\includegraphics[width=\linewidth]{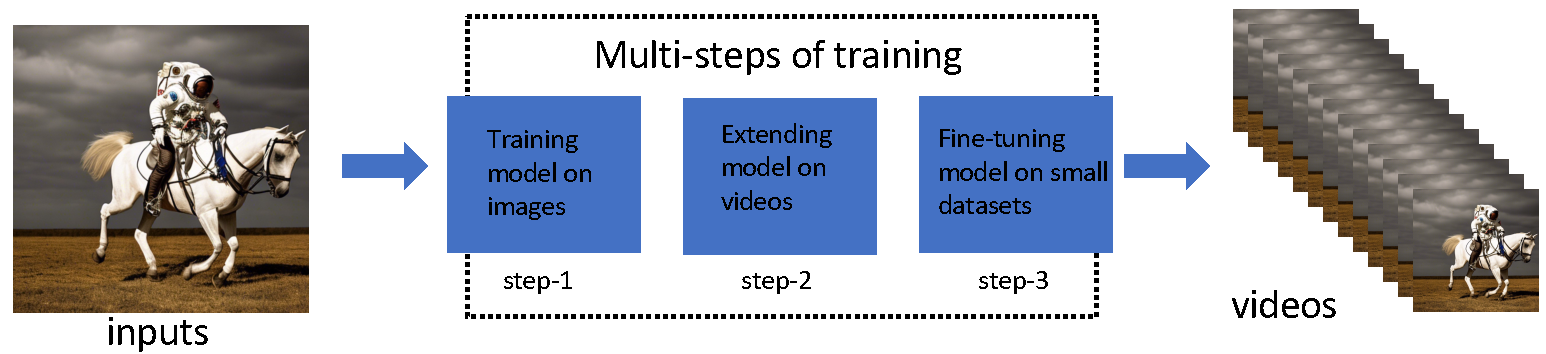}

	\caption{SVD has undergone 3 layers of training: first on images, then on video generation with temporal layers pretrained on a larger dataset, and finally fine-tuned with a smaller dataset of high-quality videos..}
	\label{fig:svd7}
\end{figure}

\subsection{Camera Motion Control}

To facilitate controlled camera motion in image-to-video generation, SVD introduces Low-Rank Adaptation (LoRA)~\cite{hu2021lora}modules. These modules are trained within the temporal attention blocks of the model and allow for the manipulation of camera motion in the generated videos.

Multi-View Generation: The paper also explores the use of the base model for multi-view generation. SVD finetunes the image-to-video model on multi-view datasets to generate consistent novel views of an object. They compare their approach with state-of-the-art methods and demonstrate superior performance.

\subsection{Formulas and Mathematical Framework}

The paper utilizes the framework of diffusion models, which is rooted in the following mathematical concepts:

Probability Flow ODE: The iterative refinement process is implemented through the numerical simulation of the Probability Flow ordinary differential equation (ODE).

Denoising Score Matching (DSM): The training of the denoiser is achieved through DSM, which minimizes the difference between the predicted and actual clean data.

Classifier-Free Guidance: A method to guide the diffusion process is given by a weighted sum of conditional and unconditional model predictions.

\begin{figure}[t]
	\centering

	\includegraphics[width=\linewidth]{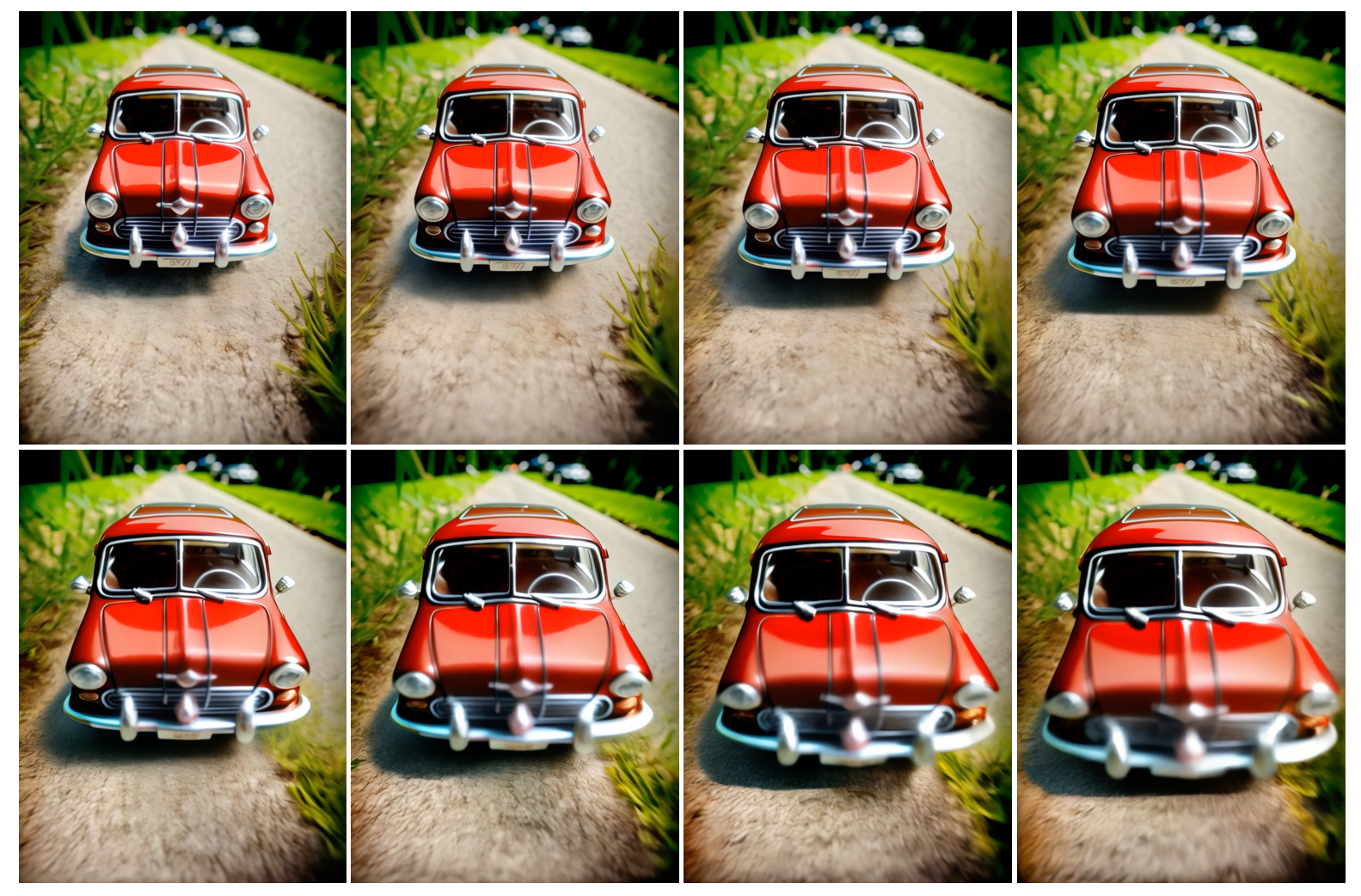}

	\caption{We shw more AI-generated image-to-video synthesis examples, effectively promoting the development of the system.}
	\label{fig:svd5}
\end{figure}

\section{Potential Design Innovation with SVD}

The integration of SVD and similar advanced generative models into the design process can offer a multitude of benefits for designers, transforming the way they approach creative tasks. Below, we outline the potential advantages, highlighting how SVD can revolutionize various aspects of design work in Fig. 3.

\subsection{Enhanced Creativity and Ideation}
\paragraph{Expanding the Bounds of Possibility}

At its core, SVD is capable of generating video content that aligns with textual or pictorial prompts provided by the designer. This capability is not just a technical feat; it's a gateway to expanding the very bounds of what designers consider possible. By inputting their initial ideas into SVD, designers can receive a plethora of video concepts that they might not have initially considered. These concepts can range from abstract visualizations of ideas to highly detailed, motion-driven sequences that were once beyond the scope of individual ideation.

\paragraph{Overcoming Creative Blocks}

Creative blocks are a common challenge in the design process. Designers may find themselves stuck, unable to visualize how an idea could be translated into a dynamic, visual medium like video. SVD can serve as a catalyst to overcome these blocks. By generating videos based on textual descriptions, it can provide fresh perspectives and unexpected directions that can reignite their creative spark.

\paragraph{Collaborative Ideation}

The collaborative nature of SVD also opens up new avenues for ideation. Designers can use it as a tool for collaborative brainstorming sessions, where team members can input ideas and immediately see them transformed into video format. This immediate visual feedback can lead to more engaged and productive ideation sessions, as team members build upon each other's ideas in real-time.

\paragraph{Rapid Visualization of Concepts}

In the past, turning a concept into a visual representation required a significant amount of time and skill, especially when it came to motion graphics and video content. With SVD, designers can rapidly visualize their concepts in video form. This rapid visualization not only speeds up the ideation process but also allows for more concepts to be explored within the same timeframe.

\subsection{Time and Cost Efficiency}
Producing high-quality video content has always been a labor of love that demands a significant investment of time and resources. Traditional video production involves numerous stages, from scripting and storyboarding to filming, editing, and post-production. Each of these stages requires specialized skills, equipment, and often, a team of professionals working for extended periods. The costs can quickly escalate, involving expenses for equipment, software, talent, and the time spent in the editing suite perfecting every shot.

With the advent of Stable Video Diffusion (SVD)~\cite{blattmann2023stable}, the landscape of video production is undergoing a transformation. SVD is a generative model that can create videos from textual or image-based prompts, effectively serving as a co-creator alongside human designers. This synergy between human creativity and artificial intelligence is a game-changer in terms of time and cost efficiency.

The acceleration of video generation with SVD is profound. Designers can input a concept or a script, and SVD rapidly generates video content that would have taken days or weeks to produce using traditional methods. This rapid generation is not just about speed; it's about unlocking the potential for designers to explore multiple iterations of their ideas quickly. The ability to generate and refine ideas swiftly leads to more innovative solutions and a more refined final product.

Moreover, the cost implications are equally significant. When the time to produce a video is drastically reduced, the associated costs with labor, equipment usage, and software licenses are also diminished. Designers can now afford to experiment with video content in ways that were previously financially prohibitive. Smaller businesses, startups, and individual creators can now access high-quality video production capabilities that were once reserved for larger entities with substantial budgets.

The efficiency gains extend to the editing and post-production process as well. With SVD, many of the manual tasks traditionally associated with video editing, such as color correction, visual effects, and motion graphics, can be automated or streamlined. This not only reduces the time spent on these tasks but also minimizes the learning curve associated with mastering complex video editing software.

Furthermore, the iterative nature of video production is greatly enhanced with SVD. Designers can receive immediate feedback on their video concepts and make adjustments on the fly. This dynamic process of creation and refinement leads to a more polished end product with less back-and-forth and fewer rounds of revisions.

Another aspect of cost efficiency is the reduction in the need for physical resources. Traditional video production often requires locations, sets, and props, all of which can be expensive and time-consuming to procure and prepare. With SVD, many of these elements can be generated digitally, reducing the reliance on physical resources and the associated costs.

The accessibility of SVD also democratizes video production, allowing a broader range of designers to create professional-grade videos. This can lead to a more diverse array of voices and ideas being represented in the video content that is produced, enriching the creative landscape.

Lastly, the environmental impact of video production can be considered. By reducing the need for physical equipment and resources, SVD aligns with sustainable production practices, minimizing the carbon footprint of video creation.

\subsection{Trend Forecasting and Innovation}

Trend forecasting in the design industry is akin to navigating a vast ocean with ever-shifting currents and unpredictable winds. Designers must have a keen eye not only on the present but also on the horizon, anticipating where the currents of style and innovation are headed next. Stable Video Diffusion (SVD) serves as a powerful tool in this navigational arsenal, providing insights into emerging trends and styles through the analysis of the videos it generates.

Moreover, SVD can generate content across various domains and styles, it exposes designers to a broad spectrum of creative expressions. This exposure can spark new ideas and inspire designers to think outside the confines of their usual creative zones. It's not just about following trends; it's about understanding the undercurrents that give rise to them and innovating from a place of informed creativity.

The innovation that stems from using SVD is twofold. Firstly, it allows designers to innovate by staying current and relevant, ensuring that their designs resonate with contemporary sensibilities. Secondly, by pushing the boundaries of what is currently popular, designers can actually influence and shape future trends. It's a symbiotic relationship where SVD serves not just as a reflector, but also as a catalyst for innovation.

In an industry that thrives on novelty and creativity, SVD provides a robust platform for designers to keep pace with the ever-evolving landscape and lead the charge. With SVD as a partner in the creative process, designers are empowered to forecast trends with greater accuracy and to innovate with a confidence born from data-driven insights.

In essence, SVD is more than a tool; it's a strategic asset that can elevate a designer's practice, fostering a deeper understanding of design dynamics and enabling a proactive approach to innovation. As the field of design continues to evolve, embracing technologies like SVD will be crucial for designers who aim to not just follow, but to forge the paths that others will follow.

\section{Conclusion}

In summary, diffusion models are transforming digital media and design, particularly in generating videos from static images. These models use denoising to create realistic video sequences, addressing challenges like consistency, motion, and style. Diffusion models offer powerful tools for creating diverse, realistic video content, reshaping visual content creation.  Applications are broad: artists animate artworks, businesses create ads, and industries produce efficient effects. Educational institutions enhance learning with dynamic materials. SVD and similar techniques boost creativity, speed up design, and cut costs. They're especially useful in animation, visual effects, and creative industries.As they evolve, their impact will grow.

\bibliographystyle{paper}
\bibliography{paper}

\end{document}